\documentclass{appolb}
\usepackage{epsfig}
\usepackage[english]{babel} 
\usepackage[square,comma,numbers,sort&compress]{natbib} 

\usepackage{overpic}

\usepackage[
   centertags, % (default) center tags vertically
   sumlimits,  %(default) Place the subscripts and superscripts of summation
   intlimits,  % Like sumlimits, but for integral symbols.
   namelimits, % (default) Like sumlimits, but for certain 'operator names' such as
]{amsmath} %
\usepackage{amssymb}

\usepackage{mciteplus}

\begin{document}
\title{Open heavy flavor at RHIC and LHC in a partonic transport model%
\thanks{Presented at Strangeness in Quark Matter 2011.}%
}
% \author{Put here the name(s) of the Author(s)
% \address{and affiliation}
% \and
% the Name(s) of other Author(s)
% \address{and their affiliation}
% }

\author{Jan Uphoff$^1$, Oliver Fochler$^1$, Zhe Xu$^2$, and Carsten Greiner$^1$
\address{$^1$ Institut f\"ur Theoretische Physik, Johann Wolfgang 
Goethe-Universit\"at Frankfurt, Max-von-Laue-Str. 1, 
D-60438 Frankfurt am Main, Germany}
\address{$^2$ Physics Department, Tsinghua University, Beijing 100084, China}
}

%\author{Jan Uphoff
%\address{Institut f\"ur Theoretische Physik, Johann Wolfgang 
%Goethe-Universit\"at Frankfurt, Max-von-Laue-Str. 1, 
%D-60438 Frankfurt am Main, Germany}
%\and
%Oliver Fochler
%\address{Institut f\"ur Theoretische Physik, Johann Wolfgang 
%Goethe-Universit\"at Frankfurt, Max-von-Laue-Str. 1, 
%D-60438 Frankfurt am Main, Germany}
%\and
%Zhe Xu
%\address{Physics Department, Tsinghua University, Beijing 100084, China}
%\and
%Carsten Greiner
%\address{Institut f\"ur Theoretische Physik, Johann Wolfgang 
%Goethe-Universit\"at Frankfurt, Max-von-Laue-Str. 1, 
%D-60438 Frankfurt am Main, Germany}
%}

\maketitle
\begin{abstract}
Heavy quarks are a unique probe to study the medium produced in ultra-relativistic heavy ion collisions. Within the partonic transport model \emph{Boltzmann approach to multi-parton scatterings} (BAMPS) the production, energy loss and elliptic flow of heavy quarks are investigated in 3+1 dimensional simulations of the quark gluon plasma evolution. With only binary interactions between heavy quarks and particles from the medium calculated within perturbative QCD, the results on elliptic flow and the nuclear modification factor are not compatible with experimental data from RHIC and LHC. 
However, if the binary cross section is multiplied with $K=4$ both the elliptic flow and the nuclear modification factor are simultaneously described at RHIC and also LHC.
Furthermore, preliminary results are presented that the implementation of radiative processes leads to a stronger suppression which agrees well with the measured nuclear modification factor at RHIC without the need of any $K$ factor.
\end{abstract}

%\PACS{PACS numbers come here}

\section{Introduction}
Several evidences \cite{Adams:2005dq,*Adcox:2004mh} show that in ultra-relativistic heavy ion collisions a unique state of matter is produced. This medium consists only of deconfined quarks and gluons and is, therefore, called quark gluon plasma. Since the plasma itself cannot be seen directly, its properties must be revealed by measuring particles produced after hadronization of the deconfined medium. Due to these indirect and complex measurements, many open questions about the properties of the produced medium remain.

Heavy quarks, that is, in this context, charm and bottom quarks, are a perfect probe to tackle many of the open questions. They are well calibrated in a sense that they are produced entirely in the early stage of the heavy ion collision due to their large mass \cite{Uphoff:2010sh} and are also tagged during hadronization due to flavor conservation. Charm (bottom) quarks hadronize to $D$ ($B$) mesons, which are, however, difficult to measure. Therefore, at RHIC one measures electrons stemming from the meson decays, which can also reveal much information about heavy quarks. However, as a disadvantage it cannot be distinguished between electrons from charm and bottom quarks. Nevertheless, at LHC for the first time it is possible to reconstruct $D$ mesons and, therefore, receive information only about charm quarks with -- in principle -- no bottom contribution.

The heavy flavor electron data from RHIC \cite{Abelev:2006db,*Adare:2006nq,Adare:2010de} and the heavy flavor electron, muon and $D$ meson data from LHC \cite{Masciocchi:2011fu,Rossi:2011nx} show that the suppression of heavy quarks is comparable to that of light quarks. From the theory perspective it was thought that radiative processes involving heavy quarks are suppressed due to the dead cone effect \cite{Dokshitzer:2001zm,Abir:2011jb}, which would imply a smaller suppression for massive particles. Elliptic flow $v_2$ measurements of particles associated with open heavy flavor also show that heavy quarks interact strongly with the other particles of the medium. Whether these observations can be explained by collisional or radiative energy loss or other effects is currently in debate \cite{Armesto:2005mz,*vanHees:2005wb,*Moore:2004tg,*Mustafa:2004dr,*Wicks:2005gt,*Adil:2006ra,*Peigne:2008nd,*Gossiaux:2008jv,*Alberico:2011zy,*Cao:2011et,*Young:2011ug,Uphoff:2011ad}.

The paper is organized as the following. After a short introduction to our model BAMPS, we compare in Sec.~\ref{flow_suppression} the results on the elliptic flow and nuclear modification factor to experimental data from RHIC and LHC. How this picture is altered by introducing radiative collisions of heavy quarks in addition to binary interactions is shown in Sec.~\ref{rad_coll}, followed by a short summary.

\section{Partonic transport model BAMPS}
For the simulation of the QGP we use the partonic transport model \emph{Boltzmann Approach to MultiParton Scatterings} (BAMPS) \cite{Xu:2004mz,*Xu:2007aa}, which describes the full space-time evolution of the QGP by solving the Boltzmann equation
%,
%\begin{equation}
%\label{boltzmann}
%\left ( \frac{\partial}{\partial t} + \frac{{\mathbf p}_i}{E_i}
%\frac{\partial}{\partial {\mathbf r}} \right )\,
%f_i({\mathbf r}, {\mathbf p}_i, t) = {\cal C}_i^{2\rightarrow 2} + {\cal C}_i^{2\leftrightarrow 3}+ \ldots  \ ,
%\end{equation}
for on-shell partons and pQCD interactions. The following processes are implemented for gluons (g) and heavy quarks (Q): $g g \rightarrow g g $, $
        g g \rightarrow g g g $, $
        g g g \rightarrow g g    $, $
        g g \rightarrow Q  \bar{Q} $, $
        Q  \bar{Q} \rightarrow g g $, $
        g Q \rightarrow g Q $ and $
        g \bar{Q} \rightarrow g \bar{Q}$.
Details of the model and the employed cross sections can be found in \cite{Xu:2004mz,*Xu:2007aa,Uphoff:2010sh,Uphoff:2011ad}. In Sec.~\ref{rad_coll} we will also discuss some preliminary results including the radiative process $g Q \rightarrow g Q g $.

\section{Elliptic flow and suppression of heavy quarks}
\label{flow_suppression}

\begin{figure}
\begin{minipage}[t]{0.49\textwidth}
\centering
\includegraphics[width=1.0\textwidth]{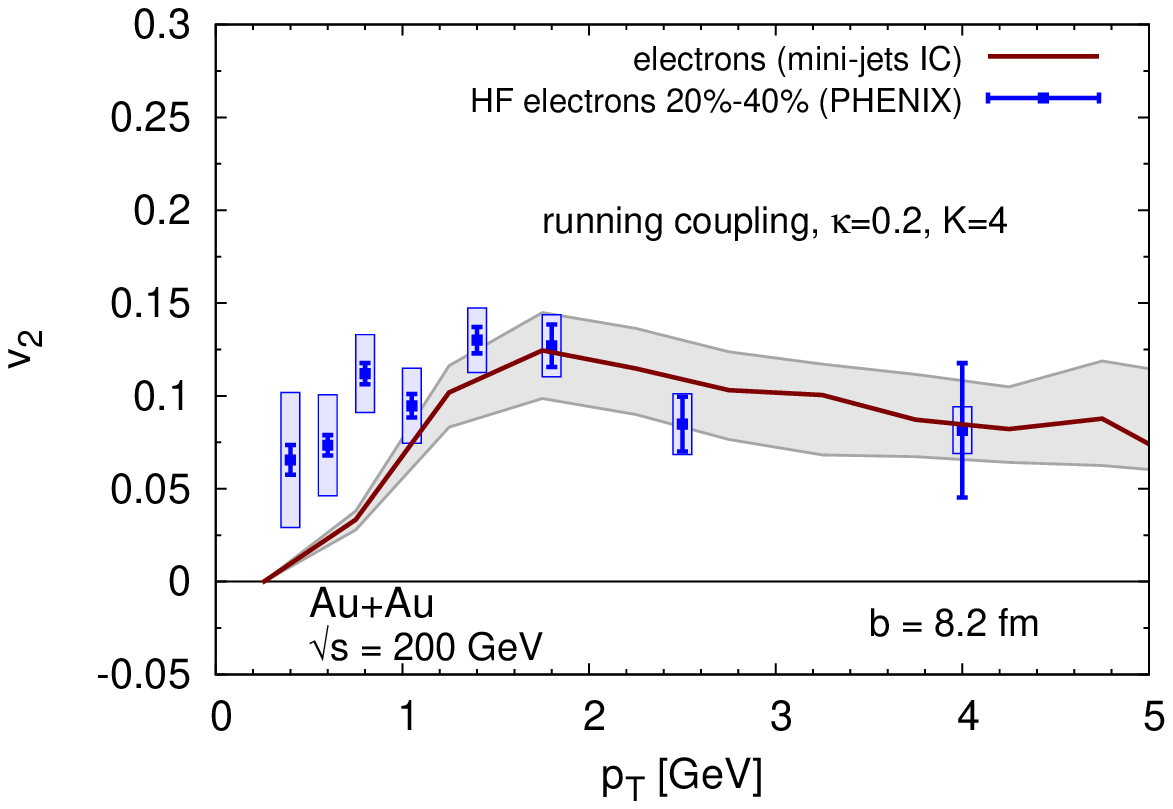}%eps
\end{minipage}
\hfill
\begin{minipage}[t]{0.49\textwidth}
\centering
\includegraphics[width=1.0\textwidth]{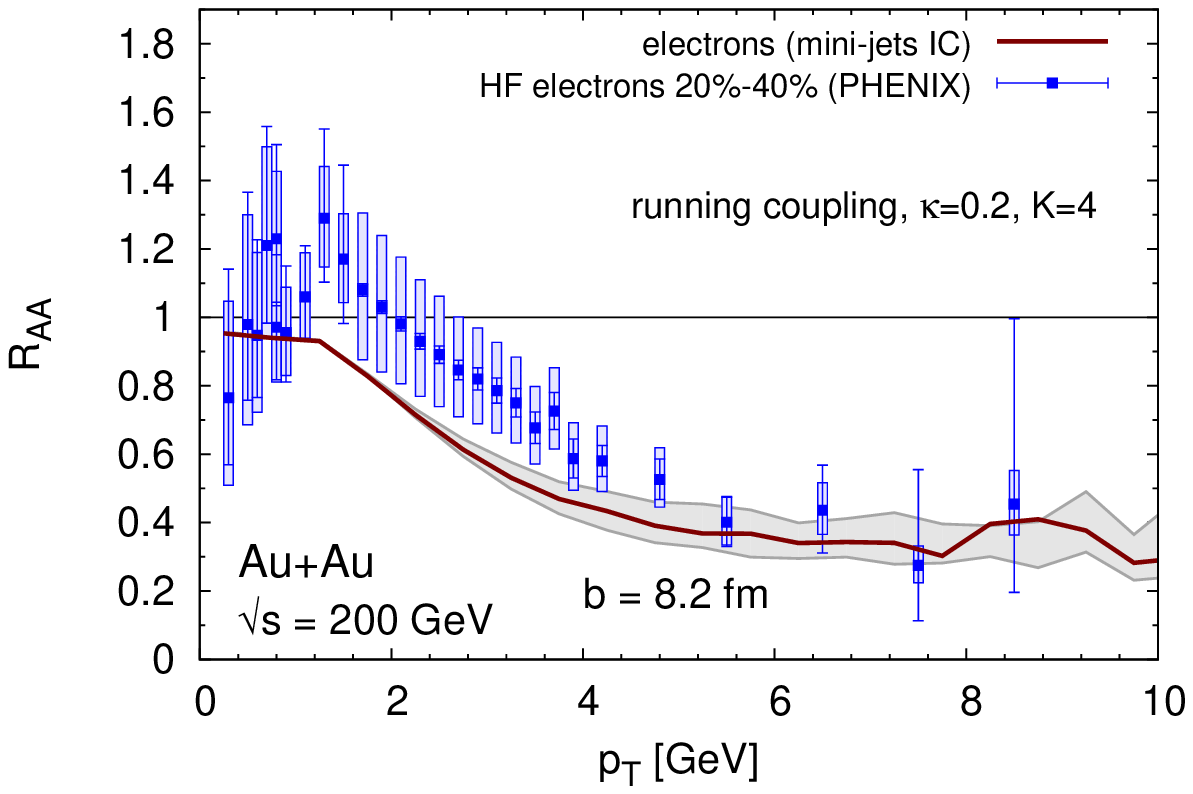}%eps
\end{minipage}
\caption{Elliptic flow $v_2$ (left) and nuclear modification factor $R_{AA}$ (right) of heavy flavor electrons for Au+Au collisions at RHIC with an impact parameter of $b=8.2 \, {\rm fm}$ together with data \cite{Adare:2010de}. The cross section of $gQ \rightarrow gQ$ is multiplied with the factor $K=4$.}
\label{fig:v2_raa_rhic}
\end{figure}

The elliptic flow $v_2$ and the nuclear modification factor $R_{AA}$ of heavy quarks are important observables. Figure~\ref{fig:v2_raa_rhic} compares our results to the heavy flavor electron data from Ref.~\cite{Adare:2010de}. 
In BAMPS the hadronization of charm (bottom) quarks to $D$ ($B$) mesons is carried out with Peterson fragmentation \cite{Peterson:1982ak,Uphoff:2011ad} and their decay to electrons is performed by \textsc{Pythia} \cite{Sjostrand:2007gs}. Since the binary pQCD cross sections are too small even with running coupling and improved Debye screening \cite{Uphoff:2010fz,*Uphoff:2010sy,*Uphoff:2010bv,*Uphoff:2011fu,Uphoff:2011ad}, those cross sections have been multiplied with a factor $K=4$ to be compatible with the data. Remarkably, both $v_2$ and $R_{AA}$ of heavy quarks can be simultaneously described with $K=4$. It has, of course, to be shown that currently missing contributions such as radiative processes can account for this additional factor. In Sec.~\ref{rad_coll} we will present a first attempt in this direction by considering the additional process $g Q \rightarrow g Q g $. 

\begin{figure}
\begin{minipage}[t]{0.49\textwidth}
\centering
\begin{overpic}[width=1.0\textwidth]{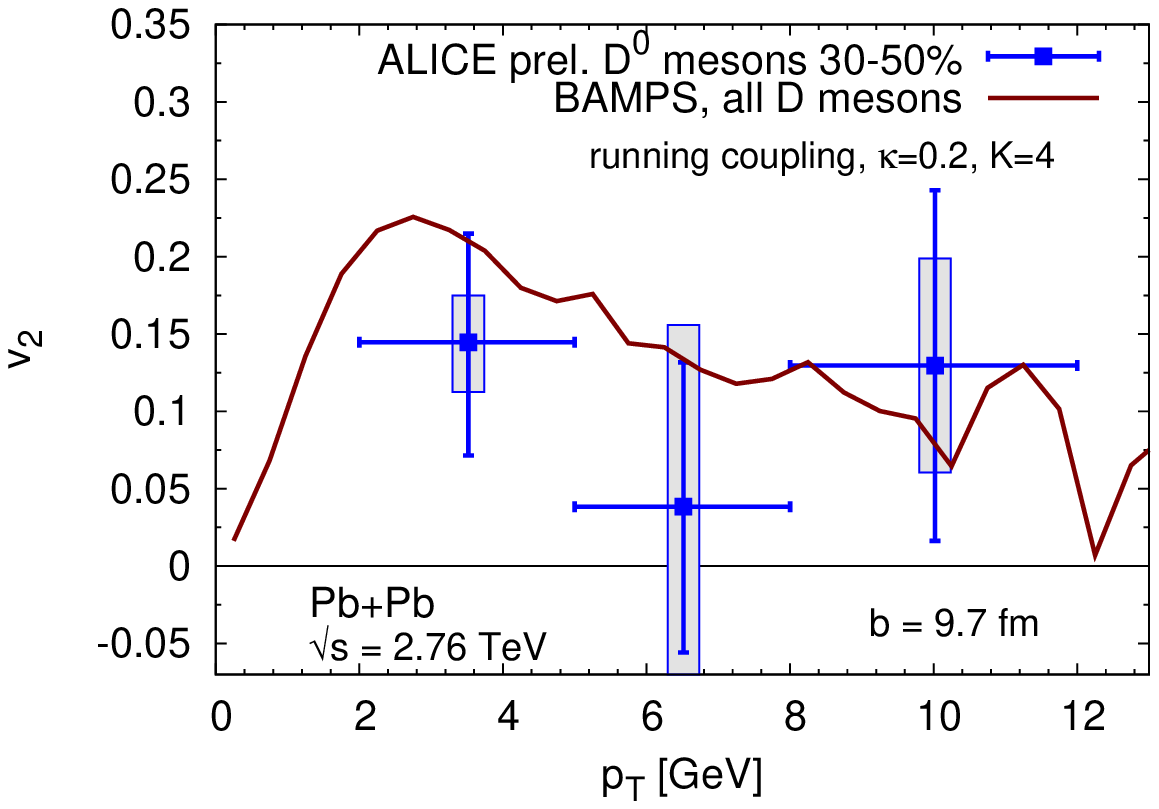}%eps
\put(48,48){\footnotesize preliminary} 
\end{overpic}
\end{minipage}
\hfill
\begin{minipage}[t]{0.49\textwidth}
\centering
\begin{overpic}[width=1.0\textwidth]{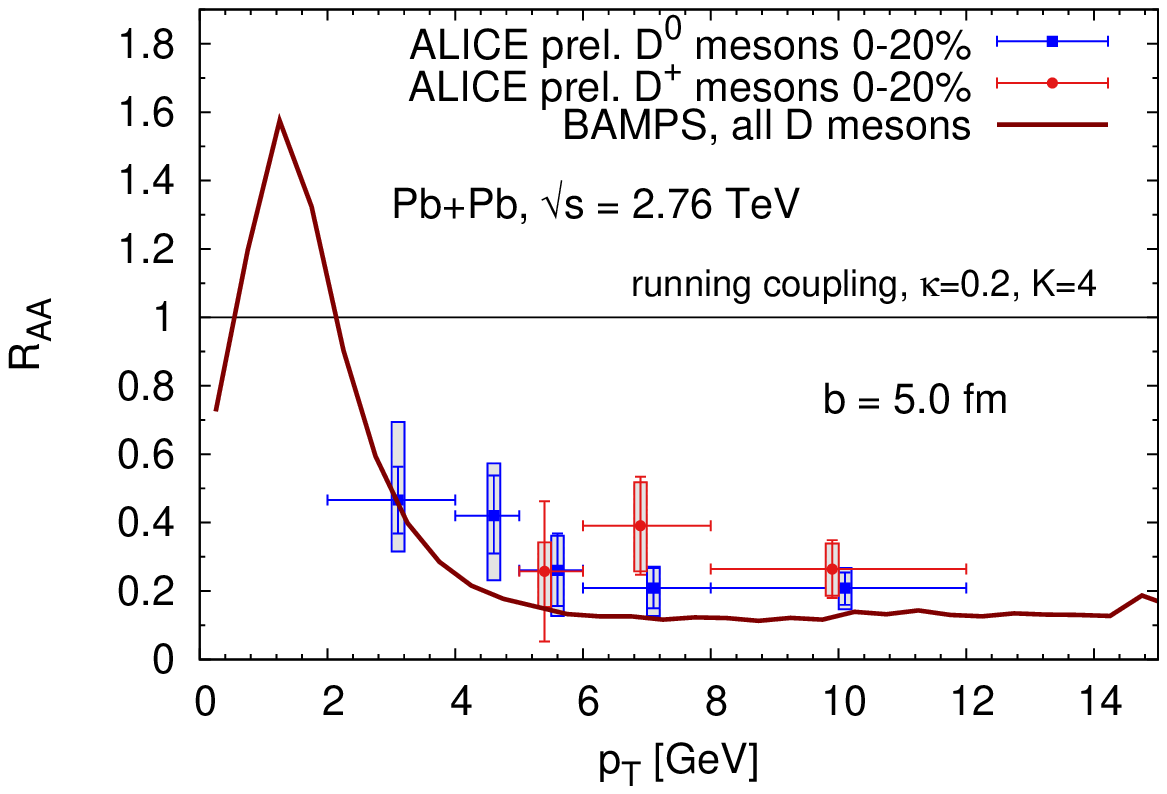}%eps
\put(63,27){\footnotesize preliminary} 
\end{overpic}
%\includegraphic%s[width=1.0\textwidth]{raa_data_dmeson_divideInitial}%eps
\end{minipage}
\caption{Preliminary BAMPS results on elliptic flow $v_2$ (left) and nuclear modification factor $R_{AA}$ (right) of $D$ mesons at Pb+Pb collisions at LHC with an impact parameter $b$ together with data \cite{Rossi:2011nx, Bianchin:2011fa}. The cross section of $gQ \rightarrow gQ$ is multiplied with the factor $K=4$.
}
\label{fig:v2_raa_d_meson_lhc}
\end{figure}

At LHC for the first time it is possible to reconstruct $D$ mesons and, therefore, distinguish between charm and bottom quarks. In Fig.~\ref{fig:v2_raa_d_meson_lhc} our results on $D$ mesons \cite{Fochler:2011en} is compared to data from ALICE.
For the same parameters, that describe the RHIC data, a good agreement is also found at LHC. In addition, muon data from charm and bottom quarks at forward rapidity is also well described, as is shown in the left hand side of Fig.~\ref{fig:raa_muon_lhc_23_rhic}.

\section{Radiative processes}
\label{rad_coll}

We calculated the Gunion-Bertsch \cite{Gunion:1981qs} matrix element for heavy quark scattering with a light quark within QCD to be \cite{Uphoff:GB}
\begin{align}
	{\left|\overline{\mathcal{M}}_{qQ \rightarrow qQg}\right|}^2 
	=12 g^2 \, 
	\left|\overline{\mathcal{M}}_{qQ \rightarrow qQ}\right|^2
	\left[ \frac{ {\bf k}_\perp}{k_\perp^2+x^2M^2} +  \frac{ {\bf q}_\perp - {\bf k}_\perp}{({\bf q}_\perp - {\bf k}_\perp)^2+x^2M^2} \right]^2 \ .
%	\nonumber
\end{align}
This is in accordance with the result obtained  with scalar QCD \cite{Gossiaux:2010yx} as well as the result in Feynman gauge \cite{Abir:2011jb} in the appropriate limit. Within the employed approximations (cf. Ref.~\cite{Gunion:1981qs}) the matrix element of the process $g Q \rightarrow g Q g $ only differs by a color factor.

\begin{figure}
\begin{minipage}[t]{0.49\textwidth}
\centering
\begin{overpic}[width=1.0\textwidth]{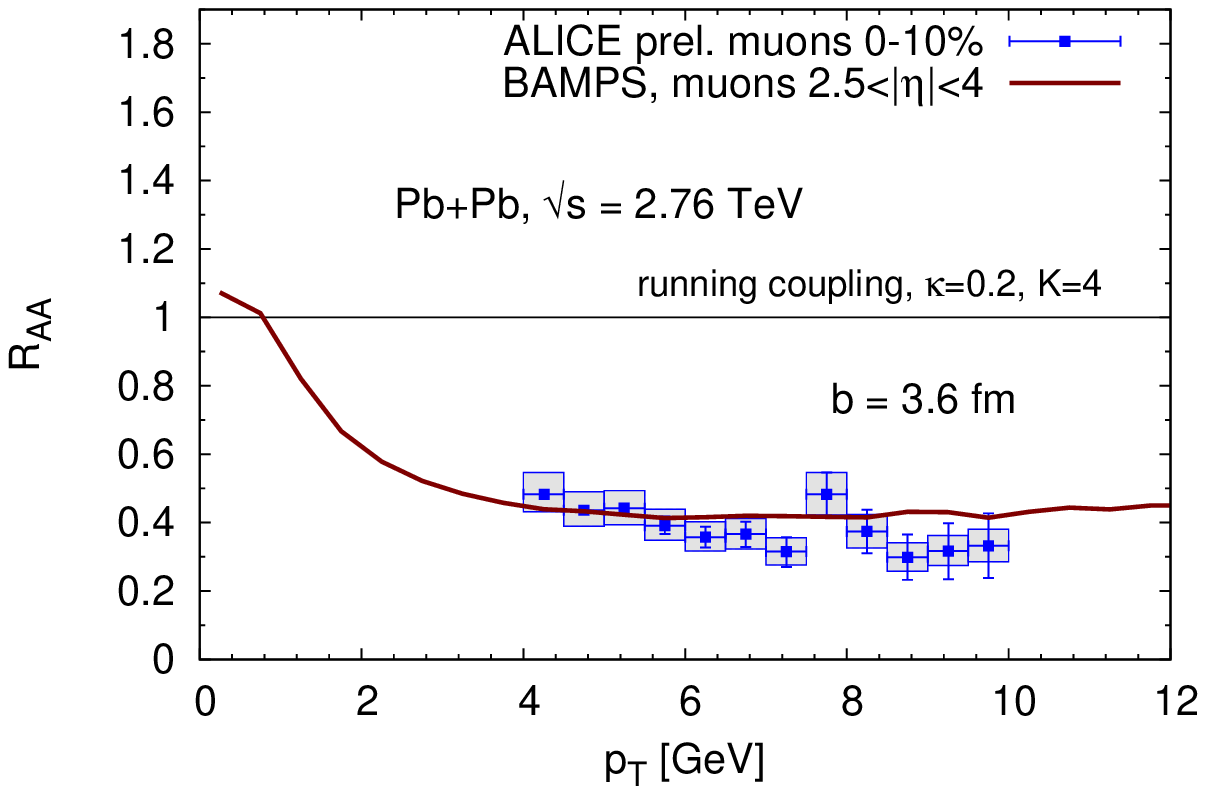}%eps
\put(19,54){\footnotesize preliminary} 
\end{overpic}
\end{minipage}
\hfill
\begin{minipage}[t]{0.49\textwidth}
\centering
\begin{overpic}[width=1.0\textwidth]{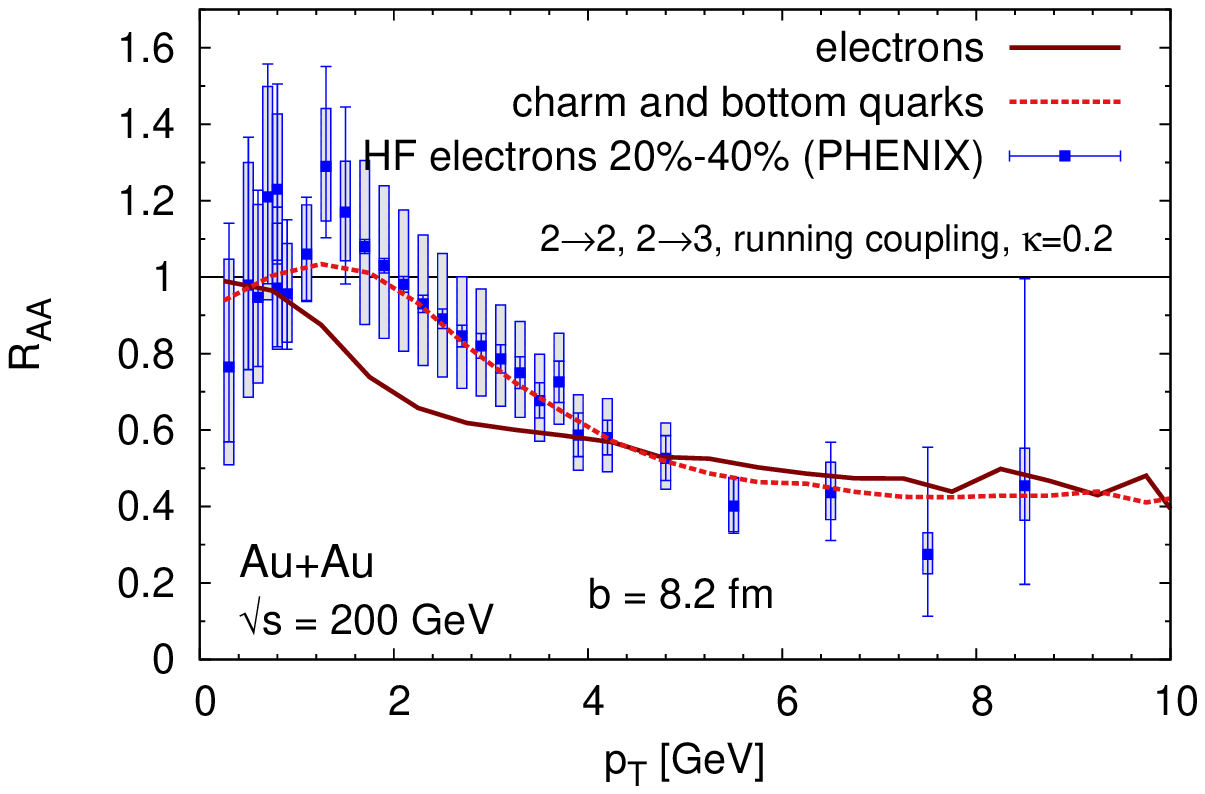}%eps
\put(19,62){\footnotesize preliminary} 
\end{overpic}
\end{minipage}
\caption{Preliminary results obtained with BAMPS. Left: $R_{AA}$ of muons at LHC with data \cite{Masciocchi:2011fu}. For heavy quarks only binary collisions are switched on, which are multiplied with $K=4$.  Right: $R_{AA}$ of heavy flavor electrons at RHIC (cf. r.h.s of Fig.~\ref{fig:v2_raa_rhic}). In addition to binary collisions, radiative processes are included. No $K$ factor is employed.}
\label{fig:raa_muon_lhc_23_rhic}
\end{figure}

In the right panel of Fig.~\ref{fig:raa_muon_lhc_23_rhic} the $R_{AA}$ of heavy flavor electrons at RHIC is shown, which is obtained with both binary and radiative processes for heavy quarks without any artificial $K$ factor. 
The curve is in good agreement with the data, in particular for large transverse momentum $p_T$ where Peterson fragmentation works best. For smaller $p_T$ coalescence might play an important role.

\section{Summary}
A full space-time simulation of the quark gluon plasma phase in heavy ion collision is presented. The elliptic flow and nuclear modification factor of heavy quarks cannot be explained by binary collisions which are calculated with pQCD, even if a running coupling and improved Debye screening is employed. To be compatible with experimental data at RHIC and LHC a $K=4$ factor is needed, which reflects that higher order contributions are currently missing in our model. First preliminary calculations with radiative contributions to heavy quark scatterings, however, look very promising since no artificial $K$ factor is needed anymore to be compatible with $R_{AA}$ data at RHIC. In the future, we will investigate the impact of the implementation of the radiative processes on heavy flavor phenomenology in more detail.

\section*{Acknowledgements}
J.U. would like to thank P. Gossiaux and J. Aichelin for insightful discussions.
The BAMPS simulations were performed at the Center for Scientific Computing of the Goethe University Frankfurt. This work was supported by the Helmholtz International Center for FAIR within the framework of the LOEWE program launched by the State of Hesse.

\bibliography{hq}

\end{document}